\documentclass[fleqn,twoside]{article}
\usepackage[headings]{espcrc2}

\readRCS
$Id: espcrc2.tex,v 1.2 2004/02/24 11:22:11 spepping Exp $
\usepackage{graphicx}
\usepackage[figuresright]{rotating}

\newcommand{\AmS}{{\protect\the\textfont2
  A\kern-.1667em\lower.5ex\hbox{M}\kern-.125emS}}
\newcommand{\Gr}{Gr\"obner }

\newcommand{\lex}{\mathop{\mathrm{lex}}\nolimits}
\newcommand{\tdeg}{\mathop{\mathrm{tdeg}}\nolimits}
\newcommand{\lt}{\mathop{\mathrm{lt}}\nolimits}
\newcommand{\lcm}{\mathop{\mathrm{lcm}}\nolimits}
\newcommand{\lc}{\mathop{\mathrm{lc}}\nolimits}

\title{\Gr Bases in Perturbative Calculations}

\author{Vladimir P.Gerdt\address[MCSD]{Laboratory of Information Technologies,
        Joint Institute for Nuclear Research,
        141980 Dubna, Russia}}

\runtitle{\Gr Bases in Perturbative Calculations}
\runauthor{V.P.Gerdt}

\begin{document}

\begin{abstract}
In this paper we outline the most general and universal algorithmic approach
to reduction of loop integrals to basic integrals. The approach is based on
computation of \Gr bases for recurrence relations derived from the integration
by parts method. In doing so we consider generic recurrence relations when
propagators have arbitrary integer powers treated as symbolic
variables (indices) for the relations.
\vspace{1pc}
\end{abstract}

\maketitle

\section{Introduction}

Since its invention in \cite{ChTk81,Tk81}, the method of integration by parts
has become the commonly used platform for reduction of loop integrals to basic
(master) integrals. The distinguishing feature of this method is rapidly increasing
volume of symbolic algebraic computation needed for manipulation with recurrence
relations (RR) when the number of internal and external lines or/and loops increases,
especially in the presence of such physical parameters as masses, external momenta
and the space-time dimension. Over the last years a number of algorithmic ideas,
approaches and software packages has been reported~\cite{Tar98,Lap00,Tar04,AnLa04}
to reduce, given a set of RR, loop integrals to a minimal set of
master integrals.

There are two main aspects of an algorithmic approach to this problem: (i) universality, i.e.
applicability to the most general form of RR and (ii) computational efficiency. Most of
practically done computations are based on the approach of Laporta~\cite{Lap00} since it deals
directly with RR and provides a rather efficient reduction procedure based on Gaussian
elimination. However, this elimination is restricted to RR with specified values of
indices. As specification takes place, one extracts only partial information from
RR. Moreover, in typical multiloop calculations it may lead to an enormous number of
intermediate RR~\cite{AnLa04} that can annihilate the computational efficiency.

Another approach suggested by Tarasov~\cite{Tar98,Tar04} appeals to potentially much more
universal algorithmic techniques based on computation of {\em \Gr bases} (GB). But for all that,
RR are transformed into partial differential equations. Such a transformation can only be done for
internal lines endowed with different masses which are independent variables
for the differential equations. Insertion of these extra variables may increase
substantially the running time and storage space required for computing GB. Furthermore, if one
considers indices of RR as (discrete) symbolic variables for the differential system, then
in the aggregate with extra independent variables (masses) this tends to increase the volume
of computation so much that becomes hardly possible for problems of interest in modern
multiloop calculations. Thus, in practice, one can treat differential systems for loop
integrals only with fixed numerical powers of propagators.

In this paper we consider RR generated by the integration by parts method as linear
finite-difference equations for loop integrals whose coefficients can (polynomially) depend on
indices and all relevant physical
parameters such as masses, scalar products of external momenta and the space-time dimension.
We present basic concepts of the GB technique (Sect.2) as the most universal algorithmic
tool for RR with symbolic indices. GB allow one to analyze
algorithmically the complete set of algebraic restrictions imposed on loop integrals by RR.
Specifically, we apply
(Sect.3) the GB technique to find a minimal set of basic (master) integrals which are
not fixed by RR, and to reduce any other loop integral to these basic ones. We illustrate
this technique by a simple example of one-loop propagator type integral taken from~\cite{Tar98}.
We also briefly discuss computational issues of the GB approach
applied to RR (Sect.4).

The concept of GB was invented by Buchberger in 1965 for systems of multivariate
commutative polynomials and then extended to systems of linear partial differential equations,
noncommutative polynomials and other algebraic structures (see, for example, book~\cite{GBA}).
Due to their algorithmic universality, GB have found numerous fruitful applications. Special program
modules for their computing in the case of commutative polynomials built-in all
modern computer algebra systems such as Maple, Mathematica and others. In addition,
Maple has special librarian functions for computing GB for differential equations, which
were used in~\cite{Tar98,Tar04}, and for noncommutative Ore algebras~\cite{Ch98}.

\section{\Gr Bases for Recurrence Relations}

Since any tensor multi-loop Feynman integral by shifting space-time dimension can be reduced to
scalar integrals~\cite{Tar96}, we shall, without the loss of generality, consider the
following scalar integrals
\begin{equation}
{\cal{I}_\nu}=\int d^dk_1\cdots d^dk_L \frac{1}{\prod_{i=1}^n P_i^{\nu_i}} \label{integral}
\end{equation}
with $n$ internal lines. Here propagators $P_i$ have symbolic exponents $\nu_i$ which we
collect together into the {\em multi index} $\nu=\{\nu_1,\nu_2,\ldots,\nu_n\}$. RR derived from
integration by parts form a homogeneous system of linear finite-difference equations in the integral
$\cal{I}_\nu$ as function of its indices. Let $\nu_i-\lambda_i$ be the minimal value of the
$i$-th index entering in the RR system. Here $\lambda_i\geq 0$ are explicitly known integers.
The set of all multi indices with nonnegative components
will be denoted by $N^n_{\geq 0}$, and, hence, $\lambda\in N^n_{\geq 0}$.

Let $\mu=\{\nu_1-\lambda_1,\nu_2-\lambda_2,\ldots,\nu_n-\lambda_n\}$ and the integral
${\cal{I}}_\mu\neq 0$ be a function of indices in $\nu$.  Then the RR are a set of
equalities $f_j=0$ including finite sums of the form
\begin{equation}
f_j=\sum_{\alpha} b_\alpha^j \, D^\alpha {\cal{I}}_\mu\,,  \qquad j=1,\ldots,p \label{fde}
\end{equation}
Here $D^\alpha=D_1^{\alpha_1}\cdots D_n^{\alpha_n}$ with
$\alpha=\{\alpha_1,\ldots,\alpha_n\}\in N^n_{\geq 0}$, and each $D_i$ is the right-shift
operator for the $i$-th index, i.e.,
$D_i{\cal{I}}_\mu={\cal{I}}_{\mu_1,\ldots,\mu_i+1,\ldots,\mu_n}$. Coefficients $b_\alpha^j$
depend polynomially on indices $\{\nu_1,\ldots,\nu_n\}$ and physical parameters, e.g., masses,
scalar products of external momenta, space-time dimension $d$. Hereafter we shall call
integrals $D^\alpha {\cal{I}}_\mu$ and sums $f_j$ in (\ref{fde}) by (difference) {\em terms}
and (linear difference) {\em polynomials},
respectively. Coefficients $b_\alpha^j$ of polynomials $f_j$ will be considered as elements
in the field $Q(\nu)$ of rational functions in indices whose coefficients in turn are parametric
rational functions.

Consider now the set of all linear polynomials
\begin{equation}
<F>=\{\ \sum_{\beta} b_\beta \, D^\beta (f)\mid f\in F,\ b_\beta\in Q(\nu)\ \}  \label{ideal}
\end{equation}
generated by the polynomial set $F=\{f_1,\ldots,f_p\}$ defined in~(\ref{fde}).

Any element $g\in \, <F>$ yields the finite-difference equation $g=0$ which is a consequence
of the initial RR, and $<F>$ accumulates all the consequences. The set $F$ of difference polynomials
is called {\em a basis} of $<F>$.

Let $\succ$ be a total order on
terms $D^\alpha {\cal{I}}_\mu$ such that for any $\alpha,\beta,\gamma \in N^n_{\geq 0}$
the following holds \\[-0.2cm]

(i) $D^\alpha {\cal{I}}_\mu \succ {\cal{I}}_\mu \Longleftrightarrow \alpha$ contains
nonzero indices, \\[-0.2cm]

(ii) $D^\gamma D^\alpha {\cal{I}}_\mu\succ D^\gamma D^\beta {\cal{I}}_\mu\Longleftrightarrow$
$D^\alpha {\cal{I}}_\mu\succ D^\beta {\cal{I}}_\mu$. \\[-0.2cm]

Such term orders are nothing else than {\em admissible term orders} in commutative polynomial
algebra~\cite{Buch98}, if one compares multi indices of the terms. As typical examples of admissible
orders we indicate ''lexicographical'' order $\succ_{\lex}$ with $\alpha \succ_{\lex} \beta$
when $\alpha_1 > \beta_1$ or when $\alpha_1 = \beta_1$ and $\alpha_2 > \beta_2$, etc., and ''total degree''
order $\alpha \succ_{\tdeg} \beta$ when $\sum \alpha_i > \sum \beta_i$ or
$\sum \alpha_i = \sum \beta_i$ and $\alpha \succ_{\lex} \beta$.

Given a term order $\succ$, one can extract the {\em leading term} $\lt(f)$ from every
difference polynomial $f$. Its coefficient is called {\em the leading coefficient}
and will be denoted by $\lc(f)$. Now we can define GB for~(\ref{ideal})
as a finite polynomial set $G$ satisfying
\begin{enumerate}
\item $<G>=<F>$. \\[-0.6cm]
\item For every $f\in \,<F>$ there is $g\in G$ and $\gamma\in N^n_{\geq 0}$ such that
$\lt(f)=D^\gamma (\lt(g))$.
\end{enumerate}
Note that conditions (i) and (ii) of admissibility for $\succ$ provide existence of GB
and termination of an algorithm for its construction exactly as they do in commutative
polynomial algebra~\cite{Buch98,CLO}. If the term order $\succ$ is considered as a certain ''simplicity
relation'' between integrals, then condition (i) means that integral ${\cal{I_\mu}}$
is ''simpler'' than that obtained from ${\cal{I_\mu}}$ by shifting up some indices.
Condition (ii) shows that such a simplicity relation between two integrals is stable under
identical shifting of their corresponding indices.

For two multi indices $\alpha$ and $\beta$, $\alpha$ is called {\em a divisor} of
$\beta$ and $\beta$ is called {\em a multiple} of $\alpha$, if
$\beta-\alpha\in N^n_{\geq 0}$. The {\em least common multiple} of $\alpha$ and $\beta$ will be denoted by
$\lcm(\alpha,\beta)$. Thus, property 2 means that the leading multi index of any
consequence of RR as element in $<F>$ has a divisor among multi indices of the leading terms in
GB. Property 1 implies that both sets of RR defined by $F$ and $G$ are fully equivalent
since any consequence of one set is a consequence of another set and vice versa. A term whose
multi index divides (is multiple of) another term is said to be a
divisor (multiple) of this term.

An important notion of the GB techniques is {\em reduction}. Polynomial $f$ is said to be
{\em reducible modulo polynomial} $g$ if $f$ has a term $u$ with a coefficient $b\in Q(\mu)$ such
that $u$ is
multiple of $\lt(g)$, i.e., $u=D^\gamma \lt(g)$ for some $\gamma \in N^n_{\geq 0}$. In this case
{\em an elementary reduction step} is given by
\begin{equation}
f\stackrel{g}{\longrightarrow} f^\star=\frac{1}{b}\,f-D^\gamma \left(\frac{1}{\lc(g)}\,g\right)\,.\label{step}
\end{equation}
If $f^\star$ in (\ref{step}) is still
reducible modulo $g$, then the second reduction step can be performed, etc., until an irreducible
polynomial obtained. By property (i) of term order $\succ$, the number of elementary
reductions is finite~\cite{CLO}. Similarly, one can {\em reduce} a polynomial $f$ {\em modulo a
finite polynomial set $F$} by doing elementary reductions of $f$ modulo individual polynomials
in $F$. The reduction process is terminated with polynomial $\bar{f}$ irreducible modulo $F$. In this
case we say that $\bar{f}$ is in the {\em normal form modulo} $F$ and write $\bar{f}=NF(f,F)$.
The above described procedure gives an algorithm for computing the normal form.

From the above definition of GB $G$ for set $F$, it follows that
$NF(f,G)=0$ for any polynomial $f$ in $<F>$. Therefore, if $G$ is known, one can algorithmically
detect the dependency of an extra recurrence relation $r$ on set $F$. It suffices to verify whether
$NF(r,G)$ vanishes or not.

It is remarkable, that there is an algorithm for construction of GB for any finite set
of linear difference polynomials and an admissible term order. This algorithm was discovered
by Buchberger in 1965 for commutative algebraic polynomials. Below we present the Buchberger
algorithm in its simplest form~\cite{Buch98,CLO}
adapted to linear finite-difference polynomials. First, we define the {\em $S$-polynomial}
$S(f,g)$ for a pair of (nonzero) difference polynomials $f$ and $g$ as follows. Let $lt(f)$ and $lt(g)$
have multi indices $\alpha$ and $\beta$, respectively, and let $\gamma=\lcm(\alpha,\beta)$. Then
$$ S(f,g)=D^{\gamma - \alpha}\left(\frac{1}{\lc(f)}\,f\right) - D^{\gamma - \beta}
 \left(\frac{1}{\lc(g)}\,g\right)\,. $$

{\em Buchberger algorithm}:
\vskip 0.1cm

Start with $G:=F$.

For a pair of polynomials $f_1,f_2\in G$:

\hspace*{0.5cm} Compute $S(f_1,f_2)$.

\hspace*{0.5cm} Compute $h:=NF(S(f_1,f_2),G)$.

\hspace*{0.5cm} If $h=0$, consider the next pair.

\hspace*{0.5cm} If $h\neq 0$, add $h$ to $G$ and iterate.
\vskip 0.1cm
\noindent
Inter-reduction of the output GB, that is, reduction of every polynomial $g\in G$
modulo $G\setminus \{g\}$, gives a {\em reduced} GB. By normalizing (dividing by)
the leading coefficients, we obtain the {\em monic} reduced GB that is uniquely
defined~\cite{Buch98,CLO} by the input polynomial set $F$ and term order $\succ$.

\section{Generic Master Integrals}

Having computed GB for the set (\ref{ideal}) generated by the initial set (\ref{fde}),
one can algorithmically, just as in commutative algebra~\cite{CLO}, find the maximal set of
difference monomials (integrals) which are independent modulo set (\ref{ideal}). In other words,
there is no nonzero polynomial in (\ref{ideal}) composed from those monomials, and any other
monomial (integral) can be reduced to (expressed by means of) these monomials. Thus, the maximal
independent set of monomials (integrals) is exactly the collection of basic integrals. They may be
called {\em generic master integrals} to emphasize their relevance to RR with symbolic indices.
The reduction of any integral to ''masters'' can also be done algorithmically (Sect.2) by means of
GB.

By keeping track of all intermediate elementary reductions~(\ref{step}), one can obtain
an explicit expression of any loop integral in terms of (generic) master integrals and
their multiples. As we defined in Sect.3, a multiple of the integral is obtained by applying
$D^\gamma$ with $\gamma \in N^n_{\geq 0}$. It is clear, however, that $D^\gamma{\cal{I}}_{\mu}$
can be evaluated only if the dependence of ${\cal{I}}_{\mu}$ on indices is explicitly known.

But how to obtain the set of master integrals? It can be easily detected from the leading
monomials of GB. Namely, all those monomials that are not multiple of any leading
monomial in GB are independent (masters). This fact is very well known in theory of
commutative \Gr bases~\cite{CLO} and apparently applicable to linear finite-difference
polynomials. Clearly, any multiple of a leading term in GB is reducible to
(expressible in terms of) a linear combination of monomials that are not multiple of any leading
term in the GB and with coefficients from $Q(\nu)$.

Though the particular form of master integrals depends on the choice of term order $\succ$,
their number is invariant on $\succ$ and fully defined by initial RR. In fact, if one
accepts the interpretation of $\succ$ as simplicity relation (Sect.2), the set of master integrals
detected from GB contains the ''simplest'' integrals\footnote{For numerically fixed values of indices
and particular integrals sometimes one uses other simplicity relations
which do not satisfy (i) and (ii)~\cite{AnLa04}.}. Usually a
''total degree'' order is preferable (cf.~\cite{Lap00,AnLa04}) from the computational point of
view since total degree GB are typically computed much faster then lexicographical
ones.

For pure illustrative purposes consider RR for the one-loop propagator type
integral ${\cal{I}}_{\nu_1\nu_2}$ taken from paper~\cite{Tar98} and put $m^2=q^2=0$ in those
relations, thus, keeping space-time dimension $d$ as the only parameter
\vskip 0.1cm
\noindent
$
\nu_1{\cal{I}}_{\nu_1-1\,\nu_2+1}-(d-\nu_1-2\nu_2){\cal{I}}_{\nu_1\nu_2}=0, \\[0.05cm]
\nu_1{\cal{I}}_{\nu_1+1\,\nu_2-1}-\nu_2{\cal{I}}_{\nu_1-1\,\nu_2+1}+(\nu_2-\nu_1){\cal{I}}_{\nu_1\nu_2}=0. \\
$ \\[-0.3cm]
Their form (\ref{fde}) for ${\cal{I}}_{\nu_1-1,\nu_2-1}\neq 0$ is given by
\vskip 0.1cm
\noindent
$[\nu_1D_2^2-(d-\nu_1-2\nu_2)]\,{\cal{I}}_{\nu_1-1,\nu_2-1}=0, \\[0.05cm]
[\nu_1D_1^2-\nu_2D_2^2+(\nu_2-\nu_1)]\,{\cal{I}}_{\nu_1-1\,\nu_2-1}=0,
$
\vskip 0.1cm
\noindent
where  $D_1$, $D_2$ are the right-shift operators for
indices $\nu_1$ and $\nu_2$, respectively.
A total degree \Gr basis at
$D_1\succ D_2$ calculated by Maple \footnote{Algebra of shift operators is a
particular case of the Ore algebra~\cite{Ch98}, and one can use it for computing GB for
RR as already mentioned in~\cite{Tar04}. However, this way of computation is
very expensive and can be applied only to small problems (Sect.4).}  gives the following GB form of RR
\vskip 0.1cm
\noindent
$[(d-\nu_1-2\nu_2)D_1D_2-\nu_1D_2^2]\,{\cal{I}}_{\nu_1-1\,\nu_2-1}=0, \\[0.05cm]
[\nu_1(d-\nu_1-2\nu_2)D_1^2+\\[0.05cm]
(\nu_1(2\nu_2-\nu_1)+\nu_2(2\nu_2-d))D^2_2]\,{\cal{I}}_{\nu_1-1\,\nu_2-1}=0,\\[0.05cm]
D^3_2\,{\cal{I}}_{\nu_1-1\,\nu_2-1}=0.
$
\vskip 0.1cm
\noindent
An immediate consequence of this GB is the equality ${\cal{I}}_{\nu_1-1\,\nu_2+2}=0$ that
follows from the last element of GB. This implies
$ D^\gamma {\cal{I}}_{\nu_1-1\,\nu_2+2}=0$ for $\gamma\in N^n_{\geq 0}\,. $
Similarly, the order $D_2\succ D_1$ gives ${\cal{I}}_{\nu_1+2\,\nu_2-1}=0$ and
$D^\gamma {\cal{I}}_{\nu_1+1\,\nu_2-1}=0$ for $\quad \gamma\in N^n_{\geq 0}$
that can also be verified by calculating the normal form
$NF(D^3_2\,{\cal{I}}_{\nu_1-1\,\nu_2-1},GB)=0$.

The leading terms in the above GB are represented by the operators
$ D_1D_2,\,D_1^2,\,D_2^3$
which generate four master integrals
${\cal{I}}_{\nu_1-1\,\nu_2-1},\,{\cal{I}}_{\nu_1\,\nu_2-1},\,{\cal{I}}_{\nu_1-1\,\nu_2},\,
{\cal{I}}_{\nu_1-1\,\nu_2+1}\,.$

\section{Computational Aspects}

First of all, it should be emphasized that the above analysis is based only on RR themselves
without any use of the ''integral structure'' of ${\cal{I}}_{\mu}$. If there are extra
restrictions which follow from this structure, they should be incorporated into RR. For example,
if a loop integral admits symmetry under some permutation of indices, one has to use RR in
the properly symmetrized form. Thus, in example of Sect.4
${\cal{I}}_{\nu_1\,\nu_2}={\cal{I}}_{\nu_2\,\nu_1}$~\cite{Tar98}, and if one adds RR with permuted
indices, then GB becomes
$D_1^2{\cal{I}}_{\nu_1-1\,\nu_2-1},\,D_1D_2{\cal{I}}_{\nu_1-1\,\nu_2-1},\,D_2^2{\cal{I}}_{\nu_1-1\,\nu_2-1}$.

It should be also noted that the GB technique is applicable to RR with specified indices. Provided
with an appropriate ordering on the integrals (priority criteria~\cite{Lap00,AnLa04}), the GB
algorithm is just Gaussian elimination~\cite{CLO}. Hence, it is conceptually identical to the
Laporta algorithm.

In the above illustrative and very simple example (Sect.3) we performed the GB calculation over the
field. Generally, however, one has to be careful in manipulation with arbitrary indices occurring
in the coefficients of difference polynomials.
In calculating GB over a coefficient field, the leading coefficients of difference polynomials
are treated generically, and division by them can be done at the intermediate steps of calculation.
GB computed this way may loose its GB properties under some specifications of parameters.
This happens when, in the course of reduction, the division by a leading coefficient was done, and this
coefficient vanishes under the specification. To avoid this trouble, one can compute GB over
a coefficient ring rather than over a field (e.g. without division). But computation over the ring
may lead to growth of intermediate coefficients and make the computation much more tedious.

As it is well-known (see~\cite{CLO} and references therein), the running time and storage space
required by the GB algorithms the GB tend to be exponential and superexponential in number of
variables (indices of RR). Besides, the presence of parameters and symbolic indices
in coefficients increases substantially the volume of computation. Therefore, to be practically
applicable to
modern multiloop calculations, one needs efficient implementation of the GB algorithms
provided by special tuning to finite-difference polynomials of type~(\ref{fde}).

To our knowledge, there are no GB packages available for multivariate finite-difference polynomials,
except those designed for the (noncommutative) skew-polynomial algebras or
Ore algebras~\cite{Ch98}. Our experimentation with the Maple package implementing the Buchberger
algorithm for skew polynomials, revealed its very low efficiency for RR with parameters.
Thus, Maple on a 2 Ghz PC with 512 Mb RAM was not able to construct (in reasonable time)
GB for the full (but still rather small) one-loop example from~\cite{Tar98} when both $m^2$
and $q^2$ are nonzero and present in RR. To our opinion, this is because of extra computational
costs caused by internal noncommutative settings of underlying built-in operations.

We find more perspective to adapt our involutive algorithms and software packages~\cite{GB98,GBY01}
to linear multivariate finite-difference polynomials. The involutive algorithms also compute GB, but
they are based on the other course of intermediate computation different from that in the Buchberger
algorithm. Our implementation of involutive algorithms already shown their higher
efficiency~\cite{GBY01} for conventional commutative polynomials. In addition, involutive algorithms
admit effective and natural parallelism, that may be crucial to manage with multiloop
calculations needed in modern higher energy physics.

\section{Acknowledgements}

This work was partially supported by the SFB/CPP-TR 9 grant and also by grants 04-01-00784 from the
Russian Foundation for
Basic Research and 2339.2003.2 from the Russian Ministry of Science and Education. I am specially
grateful O.Tarasov for longstanding useful discussions on algorithmic aspects of manipulation
with systems of differential equations arising in the loop integral reduction. I also want
to acknowledge K.Chetyrkin, M.Kalmykov and V.Smirnov for their helpful remarks.

\end{document}